%% file: HPEC_LLload_2024.tex
\documentclass[conference]{IEEEtran}
\IEEEoverridecommandlockouts
\usepackage{cite}
\usepackage{amsmath,amssymb,amsfonts}
\usepackage{algorithmic}
\usepackage{graphicx}
\usepackage{textcomp}
\usepackage{xcolor}
\def\BibTeX{{\rm B\kern-.05em{\sc i\kern-.025em b}\kern-.08em
    T\kern-.1667em\lower.7ex\hbox{E}\kern-.125emX}}
\begin{document}

\title{LLload: An Easy-to-Use HPC Utilization Tool
\thanks{DISTRIBUTION STATEMENT A. Approved for public release. Distribution is unlimited. This material is based upon work supported by the Under Secretary of Defense for Research and Engineering under Air Force Contract No. FA8702-15-D-0001. Any opinions, findings, conclusions or recommendations expressed in this material are those of the author(s) and do not necessarily reflect the views of the Under Secretary of Defense for Research and Engineering. © 2023 Massachusetts Institute of Technology. Delivered to the U.S. Government with Unlimited Rights, as defined in DFARS Part 252.227-7013 or 7014 (Feb 2014). Notwithstanding any copyright notice, U.S. Government rights in this work are defined by DFARS 252.227-7013 or DFARS 252.227-7014 as detailed above. Use of this work other than as specifically authorized by the U.S. Government may violate any copyrights that exist in this work.}
}

\author{\IEEEauthorblockN{Chansup Byun, Albert Reuther, Julie Mullen, 
LaToya Anderson, William Arcand,  Bill Bergeron,\\ 
David Bestor, Alexander Bonn, Daniel Burrill, Vijay Gadepally, Michael Houle, Matthew Hubbell, \\
Hayden Jananthan, Michael Jones, Piotr Luszczek, Peter Michaleas, Lauren Milechin, \\ 
Guillermo Morales, Andrew Prout, Antonio Rosa, Charles Yee, Jeremy Kepner 
}
\IEEEauthorblockA{
\textit{Massachusetts Institute of Technology}\\
}
}

\maketitle

\begin{abstract}
The increasing use and cost of high performance computing (HPC) requires new easy-to-use tools to enable HPC users and HPC systems engineers to transparently understand the utilization of resources. The MIT Lincoln Laboratory Supercomputing Center (LLSC) has developed a simple command, {\tt LLload}, to monitor and characterize HPC workloads. {\tt LLload} plays an important role in identifying opportunities for better utilization of compute resources. {\tt LLload} can be used to monitor jobs both programmatically and interactively.   {\tt LLload} can characterize users' jobs using various {\tt LLload} options to achieve better efficiency. This information can be used to inform the user to optimize HPC workloads and improve both CPU and GPU utilization. This includes improvements using judicious oversubscription of the computing resources. Preliminary results suggest significant improvement in GPU utilization and overall throughput performance with GPU overloading in some cases. By enabling users to observe and fix incorrect job submission and/or inappropriate execution setups, {\tt LLload} can increase the resource usage and improve the overall throughput performance. {\tt LLload} is a light-weight, easy-to-use tool for both HPC users and HPC systems engineers to monitor HPC workloads to improve system utilization and efficiency.
\end{abstract}


\input{1_introduction}

\input{1a_related_work}
\input{2_background}
\input{3_design_implementation}
\input{4_monitor_characterize}

\input{5_summary}

\section*{Acknowledgment}
The authors express their gratitude to Bob Bond, Alan Edelman, Jeffrey Gottschalk, Charles Leiserson, Kristen Malvey, Heidi Perry, Stephen Rejto, Mark Sherman and Marc Zissman for their support of this work.

\bibliographystyle{IEEEtran} 
\bibliography{HPEC_LLload_2024}


\end{document}

%% file: 1_introduction.tex
\section{Introduction}
\begin{figure*}[htbp]
   \centering
   \includegraphics[width=\textwidth]{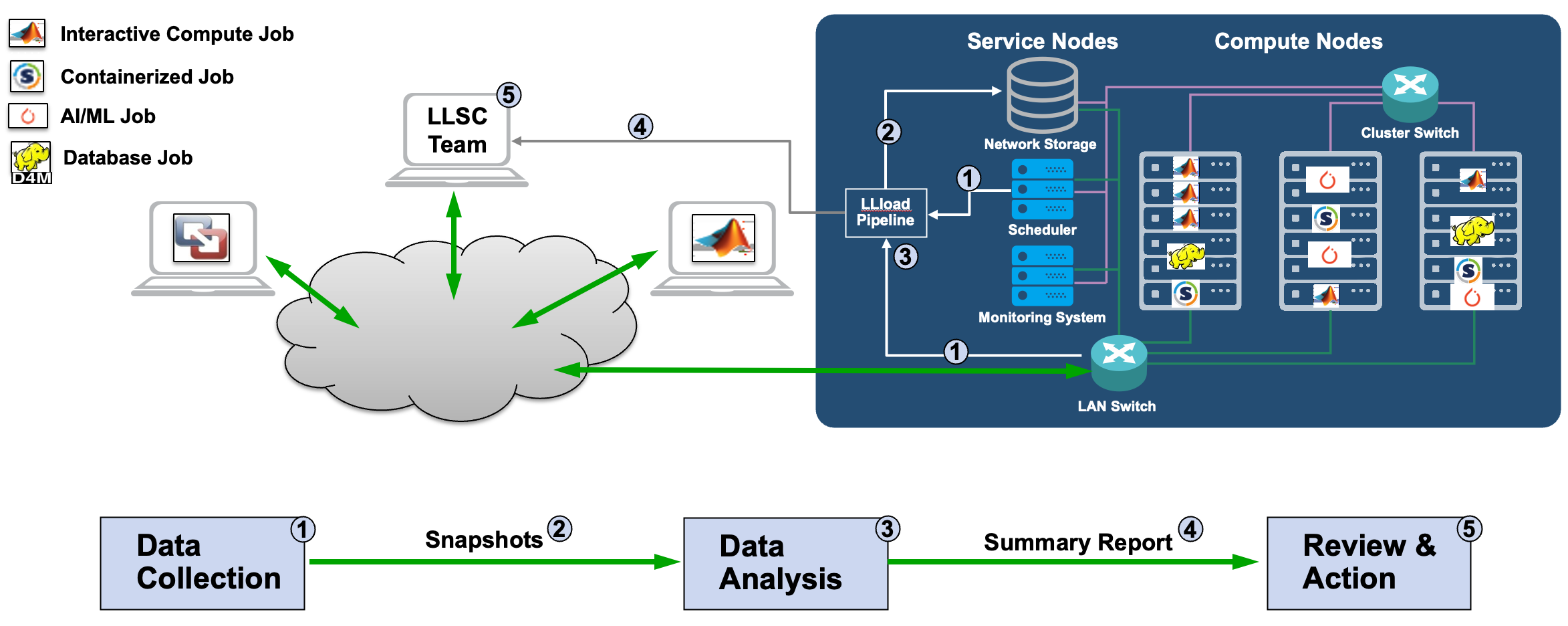}
  \caption{A schematic diagram of the LLload analysis pipeline for a supercomputing system at the MIT Lincoln Laboratory Supercomputing Center.}
  \label{LLSCISC}
\end{figure*}

A typical high performance computing (HPC) system is made of a large number of compute nodes, which are connected through fast network switches along with network storage, a scheduler and monitoring systems as shown in Figure~\ref{LLSCISC}.  
Users are connected to the HPC system via internal organization networks or external networks through login nodes and data transfer nodes.
As the system size and the number of users increase and HPC/supercomputing systems become more complex, the complexity of managing and monitoring such a big system increases as well. 
All HPC centers manage the resources and jobs on their systems with schedulers/resource managers, using automated resource allocations and policies to monitor utilization of compute and memory resources, choose the next job to run, and dispatch jobs to compute nodes for execution. 
Not surprisingly, as the user workloads are more diverse, the policies become more complex, and it becomes more challenging for users and HPC system maintainers to understand whether the resources are being used effectively and efficiently. 
Furthermore, in order to optimize system utilization, computing resources such as compute nodes and the Graphical Processing Unit (GPU), are often shared by multiple jobs which may be owned by different users. 
To add to the complexity, different types of user jobs utilize system resources in different ways. 
For instance, typical computational simulations like computational fluid dynamics (CFD) and electromagnetic simulations use a lot of system memory because they are computing changes in the state of the simulation across the large memory footprint. 
Data analysis often takes advantage of a lot of CPU capabilities, while Artificial Intelligence (AI) and Machine Learning (ML) training and inference use a lot of GPU capabilities. 
With all of these diverse workloads and complex HPC systems, most users find it quite challenging to know whether their jobs are using HPC resources effectively. 
Similarly, at HPC centers, system engineers and HPC research facilitators can become overwhelmed by the sheer volume and diversity of user applications and how the applications are using the HPC resources. 

The Lincoln Laboratory Supercomputing Center (LLSC) team saw an opportunity to develop a new, easy-to-use tool to enable HPC users and systems maintainers to quickly understand the utilization of resources.
To do this, the LLSC developed {\tt LLload}, a tool that captures a snapshot of the resources being used by active jobs on a per-user basis. 

{\tt LLload} has been deployed on LLSC systems and is successfully being used by users and the LLSC team. 
In a previous paper, we detailed the LLload design, and described how users use LLload to better understand how their applications are utilizing computational and memory resources in real-time~\cite{pearc-llload}. 
In this paper we address how the LLSC team of HPC system engineers and HPC research facilitators utilize {\tt LLload} to better understand how LLSC HPC systems are being utilized,  at the user-level and system-level. 

The {\tt LLload} tool is built from standard HPC tools and designed to provide an easy way to track the resource usage of active jobs in order to improve the performance and efficiency of HPC systems. To accomplish this, {\tt LLload} keeps track of a small set of selected metrics, namely the system CPU utilization, system GPU utilization, and memory usage metrics per node for a user or all users on a HPC system. The tool has a real-time command-line interface and can also be used to identify a list of nodes with high CPU loads or to display details of jobs and system information for a given list of nodes.


Furthermore, {\tt LLload} can be used to monitor and identify any inefficient jobs interactively or programmatically.
And by saving the LLload snapshots of all running jobs with time stamps periodically, the LLSC team can identify inefficient jobs programmatically by analyzing the collected data. 
A schematic diagram of the LLload analysis pipeline in Figure~\ref{LLSCISC} shows various stages in the pipeline including data collection, data analysis, review and action, which is implemented on a couple of supercomputing systems at the MIT Lincoln Laboratory Supercomputing Center.
We perform this analysis on a weekly basis to identify those users whose jobs had used compute resources inefficiently with low CPU and/or GPU loads.  
In addition, the weekly analysis identifies users whose jobs caused high CPU loads.  
In this paper, we discuss some background information about the LLload development, briefly recap how the tool was designed and implemented, and demonstrate how it is used to characterize users' jobs in improving their job efficiency as well as guide users in their resource requests.

%% file: 1a_related_work.tex
\section{Related Work and Tools}
Quite a number of tools have been developed to monitor how HPC systems are used.
System and job monitoring tools for large supercomputing environments focus heavily on data collection because data collection across large HPC systems without impacting user jobs is difficult. 
However, current tools are based on enterprise-oriented monitoring tools (rather than user-focused tools); are usually difficult to configure, particularly the open source tools with limited support; and/or require expert analyst knowledge to interpret and act on the data.

First, it is possible to monitor jobs using data center management tools. Data Center Infrastructure Management (DCIM) tools such as Collectd~\cite{collectd}, Telegraf~\cite{telegraf}, Ovis~\cite{ovis, Gentile17}, and InfluxDB~\cite{influxdb} require users to adhere to strict functions and syntax. 
Users enter SQL-style queries to the database and receive responses in array format.
This query/response structure requires the user to either know exactly what they are looking for or make multiple searches. Such tools are very useful for data center management staff, but due to their high learning curve and detailed feature set, they are not useful to all but the most sophisticated users and a subset of HPC system engineers.  

It is also possible to monitor jobs and users with cloud-oriented management tools. 
Prometheus~\cite{prometheus} is another open-source monitoring tool widely used in cloud environments, particularly for containerized, multi-service systems. 
However, Prometheus is not well suited for the scale and diversity of scientific workflows on HPC systems. 
An open-source tool called ruptime~\cite{ruptime} provides a command called {\tt rload} that reports CPU, GPU and their memory usage on a set of computers and can report this data on a user-by-user basis. It has low computational overhead and the command line output is easy to read, but it does not scale well to hundreds or thousands of compute nodes. 

GPU monitoring presents its own set of challenges, and therefore special monitoring tools have been developed for them. 
The gpustat tool~\cite{gpustat} is one such tools, but it is limited to Nvidia GPUs currently and not as easy to use on a HPC system since it needs to be executed locally.  
A newer version, gpustat-web \cite{gpustat-web} provides a web interface that can aggregate gpustat across multiple nodes using a remote execution of gpustat via ssh.

HPC-focused monitoring tools such as XDMoD~\cite{xdmod} and TACC Stats~\cite{taccstats} are used by many HPC centers to collect job execution data but full job execution data is not available for analysis until the job has completed. 
Ganglia \cite{ganglia} collects a wide ensemble of execution data and can display it to users during execution, but the variety of the ensemble of metrics can quickly overwhelm most novice users and can obscure the pertinent information from many HPC system engineers and research facilitators.  
Finally, the MIT Lincoln Laboratory Supercomputing Center (LLSC) has developed a 3-D game based HPC monitoring tool for HPC system engineers~\cite{Hubbell15, Bergeron21}.
This is a graphical user interface (GUI) based tool that can display the entire HPC system in a 3-D game and allows HPC system engineers and research facilitators to monitor and control the HPC system using the GUI interface. While it has features that enable multi-node job resource utilization, it was developed primarily for full-system monitoring rather than user-by-user load assessment. 

%% file: 2_background.tex
\section{Resource Sharing and Job Monitoring}

For computational performance, reliability, and security, it is important to isolate users' jobs from each other. 
The operating system kernel handles job isolation in that each process (task) gets its own resources, and HPC schedulers handle job isolation across all of the compute nodes using scheduler policies. 
By default, each core on a compute node is regarded as a single, unsharable resource, and system memory is considered a pool that is divided up among the jobs running on the node. 
Similarly GPUs can be shared among several users. This can be accomplished at a hardware level with technologies like multi-instance GPU (MIG), and and it can be accomplished at the software level by providing the same GPU ID to multiple tasks from the same user manually or with scheduler features like Slurm's CUDA Multi-Process Service (MPS).  
In general, users specify how many nodes, cores per node, system memory, and GPUs each process of their job requires for execution when they submit their jobs to the scheduler. 

Several options are available in regard to whether individual compute nodes are shared simultaneously among users. 
By default, schedulers are configured to enable the jobs of multiple users to be executed on each compute node. 
This is one policy tactic that can enable greater job throughput and utilization, but it also has some trade-offs. 
For instance, if a node fails because one of the tasks executing on it tries to use more memory than is available on the node,
all of the jobs running on that same node will fail.
This issue is particularly impactful and frustrating for users whose running jobs on the node did not cause the failure, especially those users whose jobs ran perfectly fine before.
In addition, when multiple jobs are running on the same node, it is difficult for HPC system engineers to figure out which task caused the node failure so that they can follow up with the owner of the offending job.
Another situation, which in fact happens frequently, is overloading the node by incorrectly requesting resources for each task in a job submission by the user. 
In this case, all of the jobs running on the same node will experience significant performance deterioration and slower progress of their jobs.

Schedulers do provide an option to request an exclusive run time environment for a job during the job submission; with exclusive job execution, only tasks from that given job are allowed to execute on the nodes on which that job has been dispatched to execute. 
This works for some situations, but it results in poor utilization if a user is executing many bulk synchronous parallel jobs like parameter sweeps and Monte Carlo simulations. 

Job isolation is also important for monitoring users' jobs. 
Overall, shared compute nodes also makes it difficult to monitor the resource utilization of users' jobs at the node level for the entire system. 
It is possible to aggregate the resource utilization of each process/task for each user using information in {\tt /proc/stats}, but there is a lot of data access overhead and latency involved in this process which directly affects the usability of any monitoring tool that employs this technique. 


Recently, LLSC made a significant change to their scheduling policy in order to deal with these scenarios more effectively.
LLSC has moved to a user-based whole-node scheduling policy where whole compute nodes are allocated to each user~\cite{Byun21, ByunSlurm23}.
In other words, once a user's job is dispatched to a compute node and there are unscheduled resources still available on that node, only other jobs from that same user can be scheduled on that node; jobs owned by other users cannot be scheduled on that node.
This guarantees that, at any point in time, each compute node is only executing one or more jobs/tasks owned by one and only one user. 
This makes it easier to identify which user caused the node failure and what application was responsible.
This also makes it easier to monitor individual node usage because it is now exclusively used by a single user.

When the whole-node scheduling policy was first introduced, it took some time to make the cluster system as efficient as the previous scheduling policy where all nodes are shared by all users.
The biggest issue was that short-running debugging jobs and Jupyter Notebook jobs, which use only small amount of compute resource and do not require the whole node, occupied many compute nodes and caused a shortage of available compute nodes for regular jobs that required a large number of whole nodes.
In order to resolve this issue, we deployed special partitions for servicing those debug and Jupyter Notebook jobs, which are comprised of compute nodes on which node sharing is allowed. 
To fully support the whole-node scheduling policy, LLSC updated a number of tools including LLsub, LLMapReduce~\cite{Byun16}, pMatlab/gridMatlab~\cite{Bliss07,Kepner09,Reuther04}, and pPython~\cite{Byun22, Byun23} to natively support the  whole-node scheduling policy. This includes the support of triples mode scheduling~\cite{Byun21}, where all of the tasks running on the same node are managed by a single Slurm task with an automatically generated script.

Whole-node scheduling simplifies the effort and reduces the latency of monitoring jobs so that it can be implemented as a real-time tool for users, HPC system engineers, and HPC research facilitators. 

%% file: 3_design_implementation.tex
\section{Design and Implementation}

The LLSC took a simplified, intuitive approach for job monitoring and designed LLload 
that can monitor the entire system at the user level.  
The details of design and implementation of {\tt LLload} and some of the challenges in developing {tt LLload} are described in a previous paper~\cite{pearc-llload}. 
However, for the completeness of the paper, some relavant information and additional challenges not covered in the referene paper are described below.
To keep the tool from impacting the system, especially with a large group of active users, the tool needed to 
\begin{itemize}
    \item be light-weight, 
    \item not interfere with the operation of the cluster system even under heavy user load, and
    \item use existing tools wherever possible.
\end{itemize}
Furthermore, because it is quite challenging to collect and attribute process execution metrics to the appropriate user's processes when multiple users have jobs executing on the same compute node, we needed to configure the scheduler for whole-node scheduling, which is described in the previous section.  

\begin{figure}[htbp]
   \centering
   \includegraphics[width=3.3in]{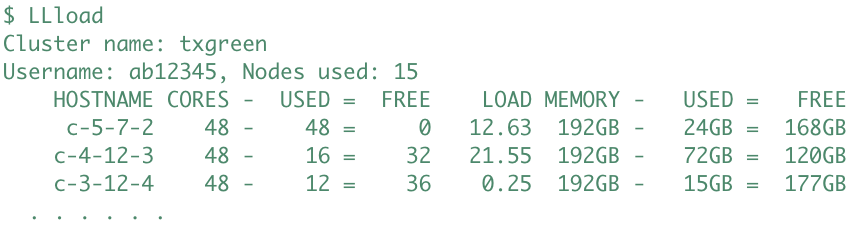}
  \caption{The default output of \texttt{LLload}, which conveys CPU utilization and system memory use.}
  \label{LLloadDefault}
\end{figure}

\begin{figure}[htbp]
   \centering
   \includegraphics[width=\linewidth]{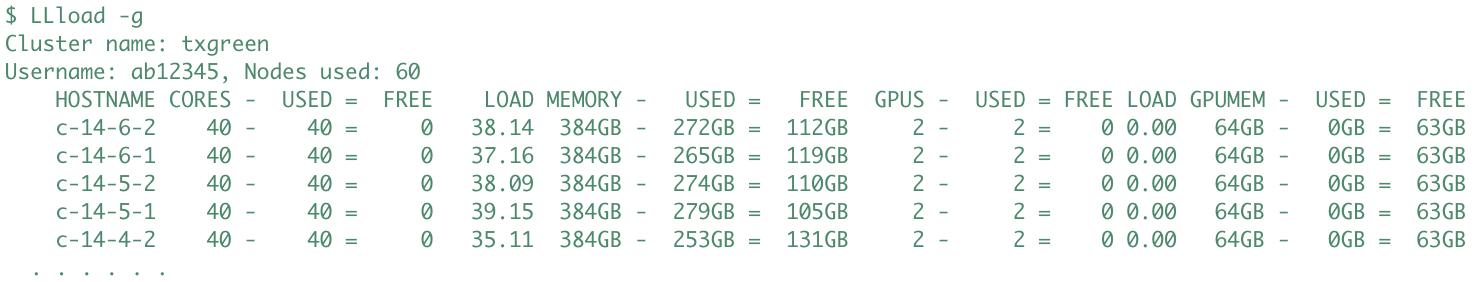}
  \caption{Typical output of \texttt{LLload} with the \texttt{-g} GPU option, which adds GPU utilization and GPU memory information along with CPU information.}
  \label{LLloadGPU}
\end{figure}

As shown in Figure~\ref{LLloadDefault} and Figure~\ref{LLloadGPU}, LLload keeps track of only a small number of metrics:  CPU (core) counts (total, used and free),  CPU memory (total, used, and free), GPU counts (total, used, and free) and GPU memory (total, used, and free) on each node.  This information is organized for each user so that we can easily identify what resources are being used by a given individual user.  

While implementing the LLload design, a few challenges surfaced. 
One is that the Slurm command {\tt sinfo} does not enable finer-grained filtering based on usernames or nodelists; it outputs all jobs that are currently executing, regardless of user. Another challenge was collecting GPU usage information. 
Both challenges are discussed in detail in the previous paper~\cite{pearc-llload}.

In addition, two additional features were added for HPC system engineers and research facilitators: the ability to view all of the users' jobs and the ability to quickly list the most overloaded compute nodes in the HPC system. 
The next several paragraphs describe these two features. 

\begin{figure}[htbp]
   \centering
   \includegraphics[width=\linewidth]{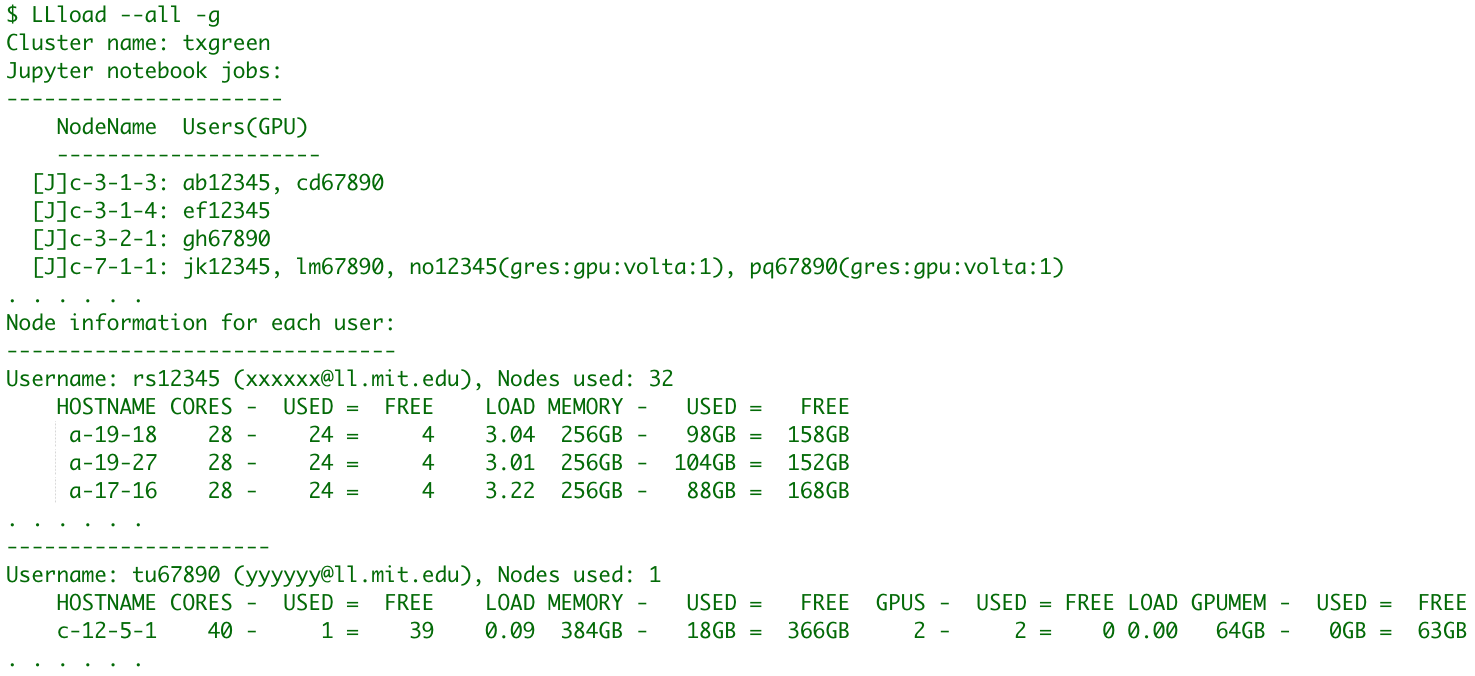}
   \caption{Typical output of \texttt{LLload} with the \texttt{--all -g} option.}
   \label{LLloadALL}
\end{figure}
The first feature added for HPC system engineers enables them to view all of the jobs currently executing on the HPC system. 
The result is the output of the {\tt LLload} with the {\tt --all} flag, which displays information on all of the compute nodes currently executing user jobs, organized by different job types and grouped by user, as shown in Figure~\ref{LLloadALL}.  
As one can see, additional information is displayed. 
First, a summary of jobs running Jupyter Notebook servers is displayed at the top. 
It displays the list of nodes and the users on each node where their Jupyter Notebook jobs are running. 
If any GPU resources are being used by Jupyter Notebook jobs, its GPU resource request is also displayed.  
Along with the usual per-user output of resources by node, the user's email information is included in the header of the user block in case there is a need to contact the user. 
The users' email addresses are maintained in separate user table and is periodically updated in order to keep up with new users, and {\tt LLload} uses this user table to obtain user email information.   
The main challenge of implementing this feature was the latency that was introduced when querying the GPU utilization of each node on which a user's jobs are executing. 
Each GPU node needs to be queried individually, and for a cleaner implementation, serially. 
This introduced a noticeable latency as {\tt LLload} collects information and displays the output to the screen. 
In the end, the team decided that this additional latency was acceptable given the value of the output that was delivered. 

It should be noted that, the {\tt --all} option is only permitted for certain LLSC staff members who have special privileges to view the information of all user jobs. 
Regular users can only see their own jobs even if the {\tt --all} option is provided.  
In addition, when extracting the GPU usage information for jobs owned by other users, it requires the remote execution on the user's behalf of a ssh command that runs {\tt nvidia-smi} as a privileged user.
Thus, LLSC implemented a special {\tt setuid} program, which is only certain LLSC staff members are allowed to execute. 
For regular users, {\tt LLload}  is only permiited to execute a remote ssh command of {\tt nvidia-smi}  on the user's behalf on each node where that user's jobs are running in order to extract their GPU usage information. 

The second feature added for HPC system engineers enables them to quickly list the most overloaded compute nodes. 
Many new users on LLSC systems are new to using HPC systems, and they do not yet understand the difference between the computational and memory resources available on their desktop/laptop versus a compute node on an HPC systems. 
This lack of understanding can easily lead to node overloading and node underutilization. 
Node overloading occurs when one or more tasks spawn many computational threads such that the total number of threads greatly exceed the number of cores available on the compute node. 
A telltale sign of this occurring is that the CPU load greatly exceeds the number of cores available on the compute node. 
For example, this situation occurs when multiple computationally intensive Python tasks are launched on the same compute node, and each of the Python tasks spawns as many threads as there are cores on the node. 
This causes a significant performance degradation because all of the Python tasks and their threads are swapping in and out of core execution cycles without getting much work done. 

\begin{figure}[htbp]
   \centering
   \includegraphics[width=\linewidth]{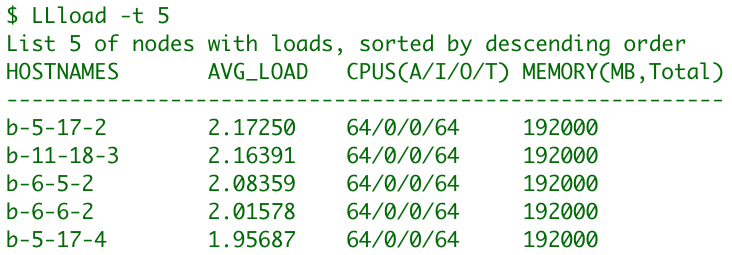}
   \caption{An output showing the top 5 compute nodes with the highest CPU loads.}
   \label{LLloadTOP}
\end{figure}

HPC system engineers frequently want to get a quick snapshot whether any compute nodes in the system are experiencing an overloaded state so that they can quickly address and mitigate the issue before it locks up the node or stalls accesses to the central parallel file systems. 
As mentioned before, the {\tt LLload} tool collects and displays the CPU load of all the compute nodes. 
This is accomplished using the Slurm {\tt sinfo} command which draws such information from the scheduler, which keeps track the 5-minute load average of each machine.
However, in this case we are more interested in identifying any compute nodes in trouble as quickly as possible.
We have implemented this feature with the {\tt -t N} option, with which {\tt LLload} displays the {\tt N} compute nodes with the highest CPU loads. 
An example of the {\tt LLload} output with the {\tt -t N} flag is shown in Figure~\ref{LLloadTOP}. 
To make the output easier to read, the CPU load is normalized by dividing the 5-minute CPU load average by the total cores available. 
This means that a fully loaded compute node will have a normalized average CPU load of around 1.0. 
If the load is greater than 1.0, it means that the node is overloaded, indicating that something is not right.
When overloaded nodes are identified, HPC system engineers can perform further investigations on those compute nodes to better understand why they are overloaded.
At this point, {\tt LLload}  does not have a similar feature to identify any compute nodes with GPU overloading.  
This feature may be implemented in the future.

Along similar lines, node underutilization occurs when not enough tasks are executing on a compute node, and compute and/or memory resources are idle because they have no computational work to execute. 
Identifying underutilized system resources is important in order to improve the system utilization and performance, but underutilization does not affect compute node stability as it does with node overloading. 
We will address node underutilization in the next Section. 

%% file: 4_monitor_characterize.tex
\section{Monitoring and Job Characterization}



In addition to the two features and their uses described in the previous Section, the {\tt LLload} tool has enabled a number of capabilities for the LLSC team. 
We will highlight two of these capabilities in this Section. 

\subsection{Weekly Analysis}
The first of these capabilities is a process with which users' job utilization trends are captured, aggregated, and analyzed on a weekly basis. 
An overview of the LLload analysis pipeline is presented in Figure~\ref{LLSCISC}.
This is particularly helpful for jobs and tasks that are underutilizing the resources on a compute node.  
Identifying jobs that have underutilized system resources is important to both users and the LLSC team because many times the cause of the underutilization can be determined and those jobs can be modified to improve system utilization. 
This, in return, increase the overall effective system performance, which means more resources are available for other users to utilize. 
Every 15 minutes, a snapshot is taken of the output of the {\tt LLload} tool and stored on the central parallel file system to archive. 
The snapshots capture the same informations as  {\tt LLload -g --all} command, but it includes a third  {\tt --tsv} flag which generates a tab-separated value (tsv) table for the output.
The tsv-format output makes it easier to ingest the data for analysis. 
Each LLSC cluster system maintains its own snapshot archive.

\begin{figure}[htbp]
   \centering
   \includegraphics[width=\linewidth]{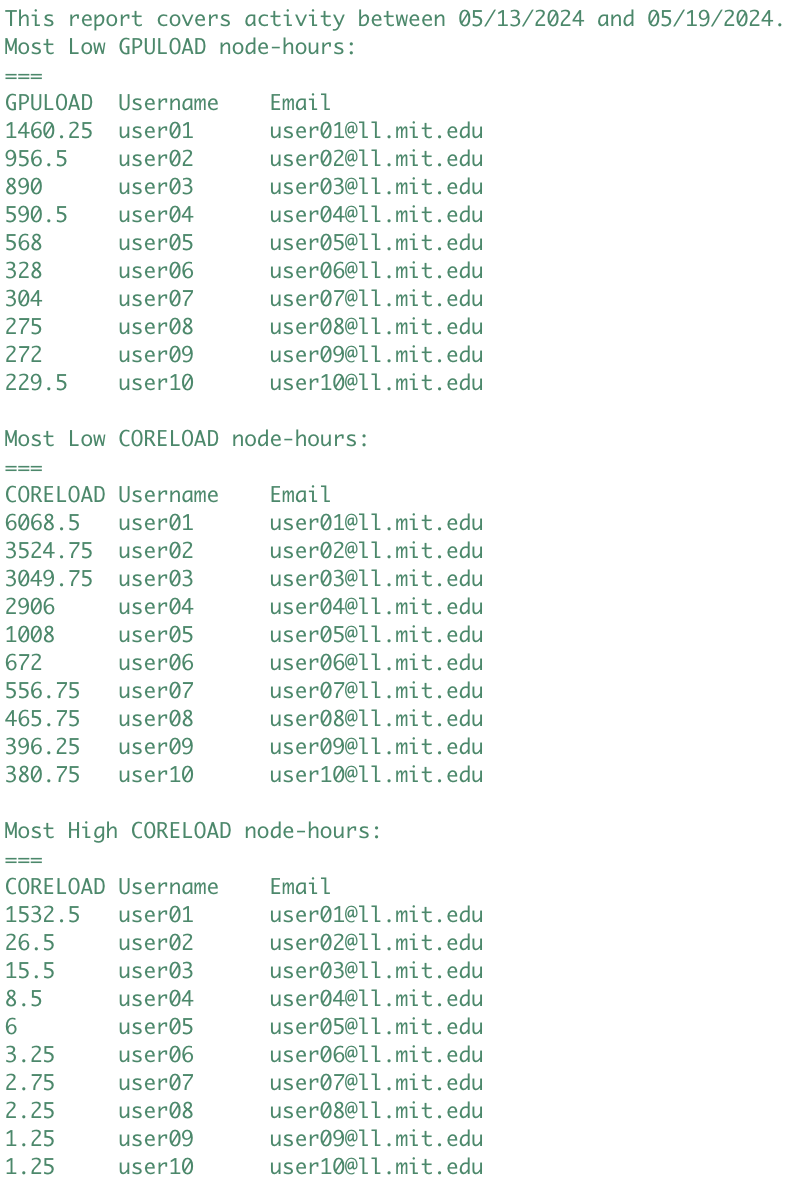}
   

  \caption{An example of the weekly report for LLSC team.}
  \label{LLloadAnalytics}
\end{figure}
We perform a weekly analysis on the archived {\tt LLload} snapshots for each system to identify the top 10 users who had executed jobs with low GPU load, low CPU load, and high CPU load.
In order to analyze the collected {\tt LLload} snapshots, we have developed Matlab/Octave code using D4M~\cite{d4m}, though similar analysis could also be accomplished with MATLAB Tables, Octave tablicious tables, or Python Pandas data frames.   
The analysis report of the 15-minute {\tt LLload} snapshots is composed of three sections: the top 10 highest aggregated node-hours of low GPU load, low CPU load, and high CPU load. 
An example of the analysis report output is shown in Figure~\ref{LLloadAnalytics}.
In this Figure, we can see that the user's information is anonymized, and a user name in one section is not necessarily the same user in the other sections.
We define GPU and CPU utilization as being underutilized when its average normalized load is below the threshold value of 0.45 ({\tt low\_threshold}). 
We define the overutilization threshold as 1+(1-{\tt low\_threshold}) where {\tt low\_threshold} is the low utilization threshold value above.
The reported GPU load and coreload values are represented in node-hour units.
The analysis code first compiles all of the instances per user that satisfy any of the low or high load conditions and then, aggregates those instances over time.
Once all of the snapshots have been computed and analyzed for the past week, the top 10 instances of each load occurrence is collected, and the results of the analysis is emailed subsequently to the LLSC staff.

\subsection{Usage Characterization}
The second of these capabilities is a process with which use the aforementioned weekly reports to initiate an email correspondence with users whose jobs have displayed high or low computational node loads. 
Each week, when the analysis report is generated and released, the LLSC team goes over each of the top 10 lists, and determines whether that user should be sent an email mentioning that our analytics had noticed certain high or low loads.
These emails include strategies and documentation links about improving the resource utilization of their jobs. 
We are judicious about sending these emails, because we do not want to send such emails to users whom we are already in the process of helping. 

We will further describe how the LLSC team uses the {\tt LLload} tool to diagnose and help users resolve resource and load issues by way of several examples. 
We have observed that experienced users usually have accrued enough knowledge to figure out how to run their jobs more efficiently on their own.
Sometimes those users report back to the LLSC team how much they have improved GPU usage with their jobs after tuning their jobs. 
However, inexperienced users may respond to the email notice asking for more information or suggestions. 
They often do not have enough experience to know what to search for in our documentation or what further techniques to try. 
We usually respond to those users with the following information:
\begin{itemize}
    \item How the low GPU usage information is generated, 
    \item Selected segments of LLload snapshots related to the user, 
    \item Selected list of jobs that caused the low GPU usage, which are identified from the snapshots, and 
    \item Suggestions such as GPU overloading to increase GPU utilization, if say a user's GPU utilization is low.
\end{itemize}

\begin{figure}[htbp]
   \centering
   \includegraphics[width=\linewidth]{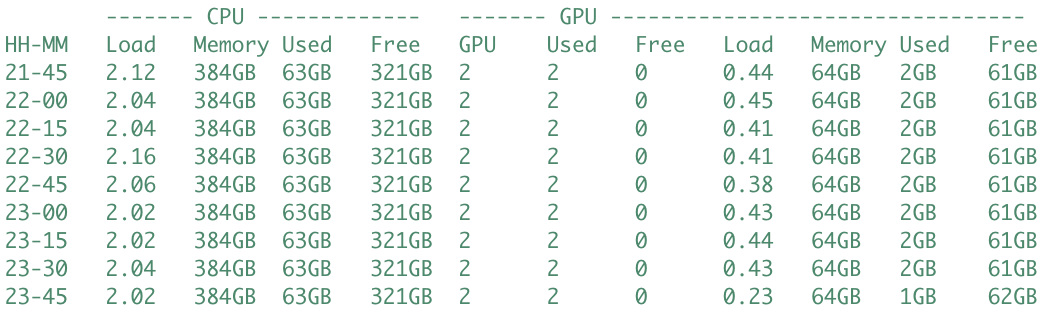}
  \caption{An example of \texttt{LLload} snapshot that highlights low GPU usage.}
  \label{LLloadSnapshot}
\end{figure}
For example, Figure~\ref{LLloadSnapshot} shows a LLload snapshot series of a user's compute node load. 
It is the simplified version of the data table that is usually sent to a user when the user has asked for additional information about the low GPU usage notification email.
This shows a series of snapshots on a compute node that summarizes both CPU and GPU usage. 
The particular job is using a modest amount of CPU memory (63 GBytes out of 384 GBytes) and minimal GPU memory (2 GBytes out of 64 GBytes).
Also, the GPU load varies between 0.23 to 0.45, which satisfies the low GPU usage threshold.
We can consider a couple of ways to improve the GPU usage for this particular case. 
One way is to increase the problem size to be processed on the GPU memory. 
For example, if the application is training an ML model, the user could increase the batch size. 
Another consideration is to assign multiple jobs to the same GPU (GPU overloading) so that the set of jobs/tasks utilize more of the GPU memory and computational capability. 
GPU overloading involves launching a parent job process through Slurm that requests access to the GPUs of the compute node. 
The parent job process subsequently launches child tasks/processes, and the parent process round-robin assigns one of the available GPUs to each of the child tasks/processes. 

\begin{figure}[htbp]
   \centering
   \includegraphics[width=\linewidth]{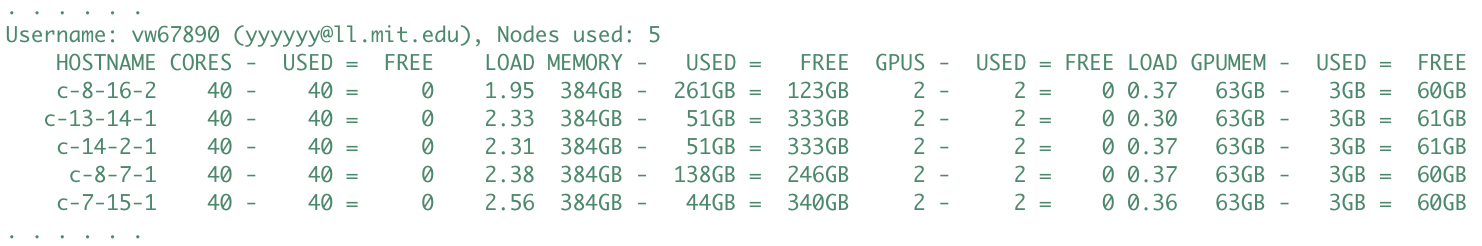}
  \caption{An example of non-optimal use of GPU nodes identified by {\tt LLload} with the {\tt --all -g} option.}
  \label{LLloadBefore}
\end{figure}
However, we have found that GPU overloading is not straight-forward to implement by most users. 
Thus, we have implemented an overloading feature for CPU and/or GPU in LLSC developed tools including LLsub and LLMapReduce so that users can enable CPU and/or GPU overloading as needed. 
With these LLSC-developed tools, we also make it easy for users to apply CPU and GPU overloading with the triples mode by increasing the number of processes per node (NPPN). 
For example, if a GPU job only needs one GPU and there are two GPUs on a node, we normally assigned two GPU jobs per node ({\tt NPPN=2}).
However, as shown in Figure~\ref{LLloadSnapshot}, based on load and memory usage, we may consider launching four ({\tt NPPN=4}) or even eight ({\tt NPPN=8}) jobs per node. In this case, the limiting factor is the GPU load.

\begin{figure}[htbp]
   \centering
   \includegraphics[width=\linewidth]{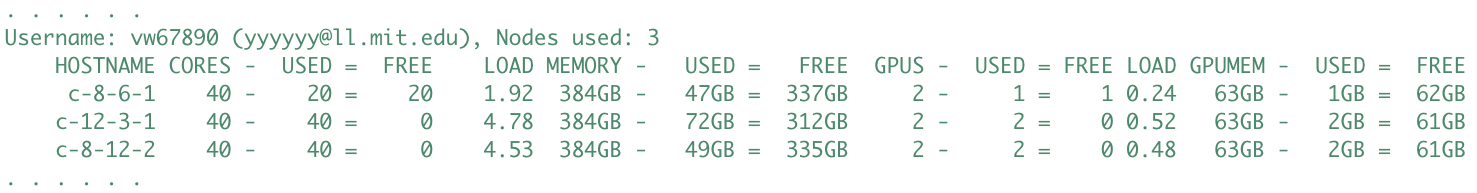}
  \caption{An example of improved usage of GPU nodes after fixing in the job submission script.}
  \label{LLloadAfter}
\end{figure}

The next example is shown in Figure~\ref{LLloadBefore} where both the GPU load and GPU memory usage are very low. 
This case shows low GPU load and low GPU memory usage and very low CPU load.
In this case, we learned that the user had requested too many CPU core resources for each job. 
As a results, although there are two GPUs on each GPU node, only one job has been dispatched per each GPU node.  
This type of mistake is common with novice HPC users.
We suggested that the user only request 20 cores per job/task and one GPU for each job. 
This change resulted in two jobs being assigned to the same node, and hence the GPU and CPU utilization increased as shown in Figure~\ref{LLloadAfter}. 
Since the user had only 5 jobs, one node ({\tt c-8-6-1}) has only one job which still shows low GPU usage.
In fact, because each job is using small amount of GPU memory in this particular case, it is possible to assign all jobs to a single node with GPU overloading, which would increase the GPU usage and free up a couple of GPU nodes for other jobs.

\begin{figure}[htbp]
   \centering
   \includegraphics[width=\linewidth]{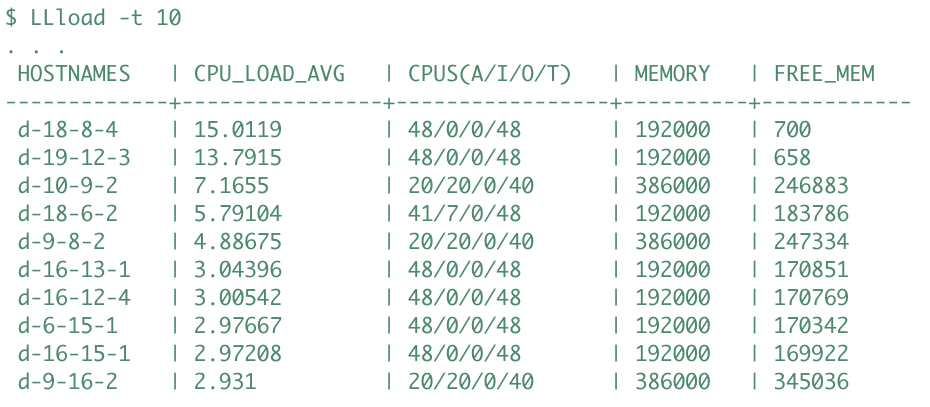}
  \caption{An example output of the \texttt{LLload -t 10} command.}
  \label{LLloadTopTen}
\end{figure}
\begin{figure*}[htbp]
   \centering
   \includegraphics[width=\textwidth]{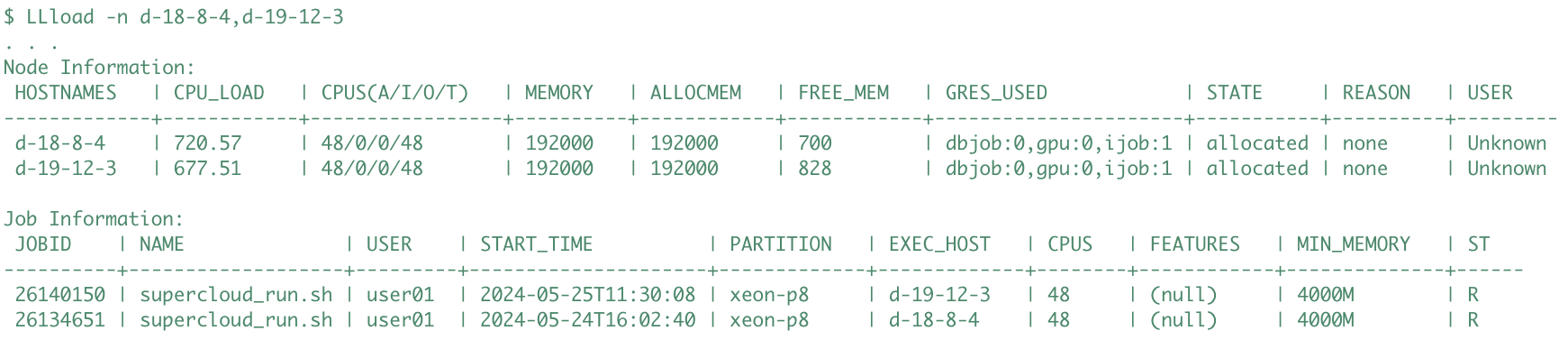}
  \caption{An example output of the \texttt{LLload -n NODELIST} command.}
  \label{LLloadNodeView}
\end{figure*}

Another case that we have often observed is that users may cause significantly high CPU load with their jobs. 
As mentioned in the previous subsection, we can identify the top 10 users with the highest aggregated node-hours of high CPU load jobs from the weekly analysis report.  
Figure~\ref{LLloadTopTen} shows an example output of the {\tt LLload -t 10} command, which shows nodes with high CPU load.
Although we identified nodes with high CPU load when the load is greater than 1.65 in the weekly report, any nodes with an average normalized CPU load above 1.0 are considered to be overloaded. 
We can see that a couple of nodes are significantly overloaded in Figure~\ref{LLloadTopTen}.
We can look at those nodes more closely in order to find out whose jobs are causing such high load by using the {\tt -n NODELIST} flag with {\tt LLload}.
Figure~\ref{LLloadNodeView} shows an example of reviewing the compute nodes identified in Figure~\ref{LLloadTopTen}.
In that Figure, both nodes are running similar jobs submitted by the same job script from the same user.
The most common cause of high CPU load is that the application is aggressively multi-threaded, meaning that it spawns as many threads as the number of physical cores or hyperthreads detected on the CPUs by the application.
This works fine when there is only one job is dispatched on each node.
This particular case shows that there is only one job is dispatched to each node but it still causes extremely high CPU load.
From interactions with the user, we learned that their application uses the Python multiprocessing package, which spawns as many threads as the number of physical threads.
However, this does not explain the cause of extremely high CPU load.
With further investigation, we found that the application executes a write statement within a loop, which issued a large number of concurrent file I/O requests.
The underlying file system client process was not able to handle these requests properly which resulted in extremely high CPU load.
We recommended that the user reduce the number of concurrent file I/Os, and the issue was resolved.

%% file: 5_summary.tex
\section{Summary}

The LLSC team has developed and deployed {\tt LLload} to monitor and characterize HPC workloads to enable better computational efficiencies on LLSC systems.
We have discussed how the {\tt LLload} tool is used by the LLSC HPC system engineers and HPC research facilitators to monitor jobs both by using LLSC-developed analytics programmatically and interactively, and what we can do with {\tt LLload} using various options to characterize HPC workloads.
Our initial study shows that we can achieve significant improvement in GPU utilization and overall throughput performance with GPU overloading in some cases but we are still investigating this area. 
In the presented examples, we have shown we can determine that some jobs may not be using the resources efficiently because their jobs are submitted incorrectly. 
By fixing incorrect job submission issue, we could significantly increase the resource usage. 
Overall, {\tt LLload}  is a light-weight, easy-to-use tool for both HPC systems engineers and users to monitor HPC workloads routinely, which is beneficial to improve system utilization and efficiency.